\documentclass[amsmath,amssymb,twocolumn, a4paper,14pt]{revtex4}
\usepackage{amsmath}
\usepackage{amssymb}
\usepackage{amsmath,amsthm,amssymb,amscd}
\usepackage{latexsym}
\usepackage{indentfirst}
\usepackage{subfigure}
\usepackage{graphicx}
\usepackage{bm}
\usepackage{multirow}

\date{\today}

\begin{document}

\title{
  \bf Microtubule length dependence of motor traffic in cells
}

\author{Yunxin Zhang}\email[Email: ]{xyz@fudan.edu.cn}
\affiliation{Laboratory of Mathematics for Nonlinear Science, Shanghai Key Laboratory for Contemporary Applied Mathematics, Centre for Computational Systems Biology, School of Mathematical Sciences, Fudan University, Shanghai 200433, China. }
\date{\today}

\begin{abstract}
In living cells, motor proteins, such as kinesin and dynein can move processively along microtubule (MT), and also detach from or attach to MT stochastically. Experiments have found that, the traffic of motor might be jammed, and various theoretical models have been designed to understand this traffic jam phenomenon. But previous studies mainly focus on motor attachment/detachment rate dependent properties. Recent experiment of Leduc {\it et al.} found that the traffic jam formation of motor protein kinesin depends also on the length of MT [Proc. Natl. Acad. Sci. U.S.A. {\bf 109}, 6100-6105 (2012)]. In this study, the MT length dependent properties of motor traffic will be analyzed. We found that MT length has one {\it critical value} $N_c$, traffic jam occurs only when MT length $N>N_c$. The jammed length of MT increases with total MT length, while the non-jammed MT length might not change monotonically with the total MT length. The critical value $N_c$ increases with motor detachment rate from MT, but decreases with motor attachment rate to MT.
\end{abstract}



\maketitle

\section{Introduction}
Motor proteins are essential for biophysical functioning of living cells \cite{Bray2001, Howard2001, Schliwa2003, Sperry2007}. Processive motors, such as kinesin and dynein, can transport organelles and vesicles along microtubule (MT) unidirectionally \cite{Vale2003, Hirokawa2003, Lipowsky2008, Stokin2005, Scholey2008, Hirokawa2010}. Experiments found that the traffic of motors along MT might be jammed \cite{Nishinari2005}, and actually, this jammed phenomenon has been predicted by various theoretical models \cite{Krug1991, Willmann2002, Parmeggiani2003, Lipowsky2006, Zhang20101, Reese2011}.

This traffic problem can be described by totally asymmetric simple exclusion process (TASEP), in which the track (MT) of motors is regarded as a one-dimensional lattice of size $N$. Each lattice site can be either empty or occupied by one motor. In the bulk, motors can hop from site $i$ to site $i+1$ with constant rate $v$ (for simplicity, $v=1$ is usually used), provided the target site is empty. At site 1, motors can enter the lattice from cell environment with density $\alpha$ , provided the site is empty. At site $N$, motors can leave the lattice into environment of density $\beta$ with rate $1-\beta$. Meanwhile, motors can attach to bulk sites of MT with rate $\omega_a$, and detach from bulk site of MT with rate $\omega_d$. The TASEP has been studied extensively \cite{Zhang20101, Parmeggiani2003, Derrida1992, Derrida1993, Kolomeisky1998}. However, previous studies about TASEP mainly focus on rates $\alpha, \beta, \omega_a, \omega_d$ dependent properties in long MT length limit, such as the existence and properties of domain wall (DW), which is defined as the interface of high motor density and low motor density  \cite{Parmeggiani2003, Raos2003, Evans2003, Reichenbach2006, Sutapa2007, Klumpp2008, Martin2009, Zhang20101}.

One surprising finding in the recent experiment of Leduc {\it et al.} is that, different from macroscale vehicle traffic, the formation of traffic jam of motor protein kinesin (Kip3) depends not only on rates $\alpha, \beta, \omega_a, \omega_d$ but also on the length of MT \cite{Leduc2012, Ross2012}. For given rate constants $\alpha, \beta, \omega_a, \omega_d$, there exists one critical value $N_c$ of MT length, DW occurs only if the MT length $N>N_c$. If there is no DW for $N\to\infty$, there will be no DW for any finite value $N$. But, even if DW occurs for $N\to\infty$, there will be no DW for MT length $N<N_c$. In cells, MT length $N$ is always finite, so it is biophysical necessary to study MT length $N$ dependent properties of motor traffic. Since the rates $\alpha, \beta, \omega_a, \omega_d$ dependent properties have been extensively analyzed \cite{Derrida1992, Derrida1993, Kolomeisky1998, Parmeggiani2003, Klumpp2008, Zhang20101}, only the MT length $N$ dependent properties of motor traffic, especially the rates $\alpha, \beta, \omega_a, \omega_d$ dependent critical MT length $N_c$,  will be discussed in this study.

\section{Method}
Let $\rho_i$ be the probability that MT site $i$ is occupied by one motor, then $\rho_i$ is governed by the following equations \cite{Zhang20101}
\begin{equation}\label{eq1}
{d\rho_i}/{dt}=v\rho_{i-1}(1-\rho_i)-v\rho_i(1-\rho_{i+1})+\omega_a
(1-\rho_i)-\omega_d \rho_i,
\end{equation}
for $2\le i\le N-1$, and at the boundaries $i=1, N$,
\begin{equation}\label{eq2}
\begin{aligned}
&{d\rho_1}/{dt}=\alpha(1-\rho_1)-v\rho_1(1-\rho_2),\cr
&{d\rho_N}/{dt}=v\rho_{N-1}(1-\rho_N)-\beta \rho_N.
\end{aligned}
\end{equation}
For kinesin Kip3 studied in \cite{Leduc2012}, $i=1$ is the minus-end of MT, and $i=N$ is the plus-end of MT.
For $N$ large enough, Eq. (\ref{eq1}) can be approximated as the following differential equation
\begin{equation}\label{eq3}
\epsilon\rho''(x)+v(2\rho(x)-1)\rho'(x)=(\Omega_a+\Omega_d)\rho(x)-\Omega_a,
\end{equation}
with boundary conditions $\rho(0)=\alpha/v$, $\rho(1)=1-\beta/v$. Where $\Omega_{a, d}=\lim N\omega_{a, d}$, and $\epsilon\to 0$ is a small parameter \cite{Parmeggiani2003}. For the special case $v=1$, the properties of motor traffic, especially the existence of DW and boundary layer (BL) of motor density $\rho(x)$, have been detailed discussed, see Tab. \ref{table1} or \cite{Zhang20101}. Let $\rho_{l\alpha}$ and $\rho_{r\beta}$ be solutions of $(2\rho(x)-1)\rho'(x)=(\Omega_a+\Omega_d)\rho(x)-\Omega_a$ but with boundary conditions $\rho_{l\alpha}(0)=\alpha$ and $\rho_{r\beta}(1)=\beta$ respectively. The results listed in Tab. \ref{table1} can be roughly summarized as follows, {\bf (I)} if $0<\alpha<0.5$ and there exists $x_0$ such that $\rho_{l\alpha}(x_0)+\rho_{r\beta}(x_0)=1$, then (a) DW occurs at $x=x_0$ if $0<x_0<1$, (b) BL occurs at left boundary $x=0$ if $x_0\le 0$, and (c) BL occurs at right boundary $x=1$ if $x_0>1$. Meanwhile, BL always occurs at right boundary $x=1$ if $0.5<\beta<1$. {\bf (II)} Otherwise, BL always occurs at left boundary $x=0$, and will also occur at right boundary $x=1$ if $0.5<\beta<1$.  

However, as mentioned before, recent experiment of Leduc {\it et al.} found that motor traffic depends also on MT length $N$,  DW occurs only when MT length $N$ large enough \cite{Leduc2012}. Therefore, the mean field results listed in Tab. \ref{table1} might not be reasonable. The MT length $N$ dependent properties of motor traffic should be discussed. The results listed in Tab. \ref{table1} indicate that, for $N\to\infty$, one necessary condition of DW formation is $0<\alpha<0.5$. So in this study, we will mainly focus on this special case, and discuss the critical MT length $N_c$.

\section{Results}
The results of long MT limit listed in Tab. \ref{table1} indicate that, there are three typical cases in which DW might occur, {\bf (1)} $0<\alpha<0.5$, $0<\beta<1/(K+1)$, where $K=\omega_a/\omega_d$, {\bf (2)} $0<\alpha<0.5$, $1/(K+1)<\beta<0.5$, and {\bf (3)} $0<\alpha<0.5$ $0.5<\beta<1$. The motor densities plotted in Figs. \ref{FigdensitysampleSKandDW}(a, b, c) show that, for any of the three cases, both the formation of DW and the ratio of jammed MT length dependent on MT length $N$. For $1/(K+1)<\beta<1$, the ratio of jammed MT length increases with MT length $N$ [see Figs. \ref{FigdensitysampleSKandDW}(a, c)], but for $0<\beta<1/(K+1)$ this ratio does not change monotonically with $N$ [see Fig. \ref{FigdensitysampleSKandDW}(b)]. The results plotted in Fig. \ref{FigdensitysampleSKandDW}(d) show that the formation of BL also depends on MT length $N$. Let $N_L$ and $N_H$ be the length of MT with low and high motor densities respectively (in calculations, low density means motor density $\rho<0.5$, and high density means $\rho\ge0.5$). The results plotted in Fig. \ref{FigMTdensity} (a-d) show that, for the cases {\bf (2) (3)}, both the high density length $N_H$ and its ratio $N_H/N$ increases with MT length $N$, the low density length $N_L$ first increases and then decreases with MT length $N$, but its ratio $N_L/N$ decreases monotonically with MT length $N$. However, for the case {\bf (1)}, neither $N_H/N$ nor $N_L/N$ changes monotonically with MT length $N$, though $N_H$ and $N_L$ increase with $N$ for almost everywhere, see Fig. \ref{FigMTdensity} (e) (f).

One meaningful question about the MT length dependent traffic of motors is that when DW will occur, provided DW occurs in large $N$ limit. Let $N_c$ be the {\it critical value} of MT length, i.e. DW occures iff MT length $N>N_c$. The curves plotted in Fig. \ref{FigCritical} show that, $N_c$ decreases with motor attachment rate $\alpha$ to the first site of MT, but increases with detachment rate $\beta$ from the last site of MT. Meanwhile, $N_c$ increases with attachment rate $\omega_a$ and deceases with detachment rate $\omega_d$. This means that, motors will be more likely to be jammed for high motor attachment rates ($\alpha$ and $\omega_d$) and low motor detachment rates ($\beta$ and $\omega_d$).

Finally, the curves plotted in Fig. \ref{FigPropOfV} show that both the non-jammed MT length $N_L$ and its ratio $N_L/N$ increase with motor velocity $v$, while the jammed MT length $N_H$ and its ratio $N_H/N$ decrease with $v$. So slow motors are more likely to be jammed. One can easily deduce that the critical length $N_c$ of MT increases with motor velocity $v$, which means that long MT is needed to block the traffic of fast motors.

\section{Conclusions}
Since the motor traffic along microtubule (MT) depends not only on motor attachment/detachment rates \cite{Zhang20101}, but also on MT length \cite{Leduc2012}, the previous results obtained by assuming MT length $N$ tends to infinity might be misleading. Therefore, in this study, the MT length dependent properties of motor traffic are studied. We found that, for given attachment and detachment rates the MT length has one critical value $N_c$, domain wall occurs only when MT length $N>N_c$. This critical length $N_c$ increases with detachment rate but decreases with attachment rate. The jammed MT length $N_H$ increases with MT length $N$, but depending on the motor detachment rate $\beta$ from the last MT site, the non-jammed MT length $N_L$ might not change monotonically with $N$. Meanwhile, for not too small motor detachment rate, $\beta>1/(K+1)$, the ratio of jammed MT length increases monotonically with MT length $N$. Calculations also show that, both the non-jammed MT length $N_L$ and its ratio $N_L/N$ increase with motor velocity $v$. In a words, the motor traffic will be more likely to be jammed for slow motors, or motors with high attachment rate to MT and low detachment rate from MT. The results in this study are helpful to understand corresponding experimental observations \cite{Leduc2012}. This study also tells us that we should be more careful to draw conclusions from continuous theoretical models for microscopic biophysical process.

\begin{acknowledgements}
This study is funded by the Natural Science Foundation of Shanghai (under Grant No. 11ZR1403700).
\end{acknowledgements}


\begin{thebibliography}{28}
\expandafter\ifx\csname natexlab\endcsname\relax\def\natexlab#1{#1}\fi
\expandafter\ifx\csname bibnamefont\endcsname\relax
  \def\bibnamefont#1{#1}\fi
\expandafter\ifx\csname bibfnamefont\endcsname\relax
  \def\bibfnamefont#1{#1}\fi
\expandafter\ifx\csname citenamefont\endcsname\relax
  \def\citenamefont#1{#1}\fi
\expandafter\ifx\csname url\endcsname\relax
  \def\url#1{\texttt{#1}}\fi
\expandafter\ifx\csname urlprefix\endcsname\relax\def\urlprefix{URL }\fi
\providecommand{\bibinfo}[2]{#2}
\providecommand{\eprint}[2][]{\url{#2}}

\bibitem[{\citenamefont{Bray}(2001)}]{Bray2001}
\bibinfo{author}{\bibfnamefont{D.}~\bibnamefont{Bray}},
  \emph{\bibinfo{title}{Cell movements: from molecules to motility, 2nd Edn}}
  (\bibinfo{publisher}{Garland, New York}, \bibinfo{year}{2001}).

\bibitem[{\citenamefont{Howard}(2001)}]{Howard2001}
\bibinfo{author}{\bibfnamefont{J.}~\bibnamefont{Howard}},
  \emph{\bibinfo{title}{Mechanics of Motor Proteins and the Cytoskeleton}}
  (\bibinfo{publisher}{Sinauer Associates and Sunderland, MA},
  \bibinfo{year}{2001}).

\bibitem[{\citenamefont{Schliwa}(2003)}]{Schliwa2003}
\bibinfo{author}{\bibfnamefont{M.}~\bibnamefont{Schliwa}},
  \emph{\bibinfo{title}{Molecular Motors}} (\bibinfo{publisher}{Wiley-Vch,
  Weinheim}, \bibinfo{year}{2003}).

\bibitem[{\citenamefont{Sperry}(2007)}]{Sperry2007}
\bibinfo{author}{\bibfnamefont{A.~O.} \bibnamefont{Sperry}},
  \emph{\bibinfo{title}{Molecular Motors: Methods and Protocols (Methods in
  Molecular Biology Vol 392)}} (\bibinfo{publisher}{Humana Press Inc., Totowa,
  New Jersey}, \bibinfo{year}{2007}).

\bibitem[{\citenamefont{Vale}(2003)}]{Vale2003}
\bibinfo{author}{\bibfnamefont{R.~D.} \bibnamefont{Vale}},
  \bibinfo{journal}{Cell} \textbf{\bibinfo{volume}{112}}, \bibinfo{pages}{467}
  (\bibinfo{year}{2003}).

\bibitem[{\citenamefont{Hirokawa and Takemura}(2003)}]{Hirokawa2003}
\bibinfo{author}{\bibfnamefont{N.}~\bibnamefont{Hirokawa}} \bibnamefont{and}
  \bibinfo{author}{\bibfnamefont{R.}~\bibnamefont{Takemura}},
  \bibinfo{journal}{TRENDS in Biochemical Sciences}
  \textbf{\bibinfo{volume}{28}}, \bibinfo{pages}{558} (\bibinfo{year}{2003}).

\bibitem[{\citenamefont{M\"{u}ller et~al.}(2008)\citenamefont{M\"{u}ller,
  Klumpp, and Lipowsky}}]{Lipowsky2008}
\bibinfo{author}{\bibfnamefont{M.~J.~I.} \bibnamefont{M\"{u}ller}},
  \bibinfo{author}{\bibfnamefont{S.}~\bibnamefont{Klumpp}}, \bibnamefont{and}
  \bibinfo{author}{\bibfnamefont{R.}~\bibnamefont{Lipowsky}},
  \bibinfo{journal}{Proc. Natl. Acad. Sci. USA} \textbf{\bibinfo{volume}{105}},
  \bibinfo{pages}{4609} (\bibinfo{year}{2008}).

\bibitem[{\citenamefont{Stokin et~al.}(2005)\citenamefont{Stokin, Lillo,
  Falzone, Brusch, Rockenstein, Mount, Raman, Davies, Masliah, Williams
  et~al.}}]{Stokin2005}
\bibinfo{author}{\bibfnamefont{G.~B.} \bibnamefont{Stokin}},
  \bibinfo{author}{\bibfnamefont{C.}~\bibnamefont{Lillo}},
  \bibinfo{author}{\bibfnamefont{T.~L.} \bibnamefont{Falzone}},
  \bibinfo{author}{\bibfnamefont{R.~G.} \bibnamefont{Brusch}},
  \bibinfo{author}{\bibfnamefont{E.}~\bibnamefont{Rockenstein}},
  \bibinfo{author}{\bibfnamefont{S.~L.} \bibnamefont{Mount}},
  \bibinfo{author}{\bibfnamefont{R.}~\bibnamefont{Raman}},
  \bibinfo{author}{\bibfnamefont{P.}~\bibnamefont{Davies}},
  \bibinfo{author}{\bibfnamefont{E.}~\bibnamefont{Masliah}},
  \bibinfo{author}{\bibfnamefont{D.~S.} \bibnamefont{Williams}},
  \bibnamefont{et~al.}, \bibinfo{journal}{Neuron}
  \textbf{\bibinfo{volume}{307}}, \bibinfo{pages}{1282} (\bibinfo{year}{2005}).

\bibitem[{\citenamefont{Scholey}(2008)}]{Scholey2008}
\bibinfo{author}{\bibfnamefont{J.~M.} \bibnamefont{Scholey}},
  \bibinfo{journal}{J. Cell Biol.} \textbf{\bibinfo{volume}{180}},
  \bibinfo{pages}{23} (\bibinfo{year}{2008}).

\bibitem[{\citenamefont{Hirokawa et~al.}(2010)\citenamefont{Hirokawa, Niwa, and
  Tanaka}}]{Hirokawa2010}
\bibinfo{author}{\bibfnamefont{N.}~\bibnamefont{Hirokawa}},
  \bibinfo{author}{\bibfnamefont{S.}~\bibnamefont{Niwa}}, \bibnamefont{and}
  \bibinfo{author}{\bibfnamefont{Y.}~\bibnamefont{Tanaka}},
  \bibinfo{journal}{Neuron} \textbf{\bibinfo{volume}{68}}, \bibinfo{pages}{610}
  (\bibinfo{year}{2010}).

\bibitem[{\citenamefont{Nishinari et~al.}(2005)\citenamefont{Nishinari, Okada,
  Schadschneider, and Chowdhury}}]{Nishinari2005}
\bibinfo{author}{\bibfnamefont{K.}~\bibnamefont{Nishinari}},
  \bibinfo{author}{\bibfnamefont{Y.}~\bibnamefont{Okada}},
  \bibinfo{author}{\bibfnamefont{A.}~\bibnamefont{Schadschneider}},
  \bibnamefont{and}
  \bibinfo{author}{\bibfnamefont{D.}~\bibnamefont{Chowdhury}},
  \bibinfo{journal}{Phys.Rev. Lett.} \textbf{\bibinfo{volume}{95}},
  \bibinfo{pages}{118101} (\bibinfo{year}{2005}).

\bibitem[{\citenamefont{Krug}(1991)}]{Krug1991}
\bibinfo{author}{\bibfnamefont{J.}~\bibnamefont{Krug}}, \bibinfo{journal}{Phys.
  Rev. Lett.} \textbf{\bibinfo{volume}{67}}, \bibinfo{pages}{1882}
  (\bibinfo{year}{1991}).

\bibitem[{\citenamefont{Willmann et~al.}(2002)\citenamefont{Willmann, Schutz,
  and Challet}}]{Willmann2002}
\bibinfo{author}{\bibfnamefont{R.~D.} \bibnamefont{Willmann}},
  \bibinfo{author}{\bibfnamefont{G.~M.} \bibnamefont{Schutz}},
  \bibnamefont{and} \bibinfo{author}{\bibfnamefont{D.}~\bibnamefont{Challet}},
  \bibinfo{journal}{Physica A} \textbf{\bibinfo{volume}{316}},
  \bibinfo{pages}{430} (\bibinfo{year}{2002}).

\bibitem[{\citenamefont{Parmeggiani et~al.}(2003)\citenamefont{Parmeggiani,
  Franosch, and Frey}}]{Parmeggiani2003}
\bibinfo{author}{\bibfnamefont{A.}~\bibnamefont{Parmeggiani}},
  \bibinfo{author}{\bibfnamefont{T.}~\bibnamefont{Franosch}}, \bibnamefont{and}
  \bibinfo{author}{\bibfnamefont{E.}~\bibnamefont{Frey}},
  \bibinfo{journal}{Physical Review Letters} \textbf{\bibinfo{volume}{90}},
  \bibinfo{pages}{086601} (\bibinfo{year}{2003}).

\bibitem[{\citenamefont{Lipowsky et~al.}(2006)\citenamefont{Lipowsky, Chai,
  Klumpp, Liepelt, and Muller}}]{Lipowsky2006}
\bibinfo{author}{\bibfnamefont{R.}~\bibnamefont{Lipowsky}},
  \bibinfo{author}{\bibfnamefont{Y.}~\bibnamefont{Chai}},
  \bibinfo{author}{\bibfnamefont{S.}~\bibnamefont{Klumpp}},
  \bibinfo{author}{\bibfnamefont{S.}~\bibnamefont{Liepelt}}, \bibnamefont{and}
  \bibinfo{author}{\bibfnamefont{M.~J.~I.} \bibnamefont{Muller}},
  \bibinfo{journal}{Physica A} \textbf{\bibinfo{volume}{372}},
  \bibinfo{pages}{34} (\bibinfo{year}{2006}).

\bibitem[{\citenamefont{Zhang}(2010)}]{Zhang20101}
\bibinfo{author}{\bibfnamefont{Y.}~\bibnamefont{Zhang}},
  \bibinfo{journal}{Chinese Journal of Physics} \textbf{\bibinfo{volume}{48}},
  \bibinfo{pages}{607} (\bibinfo{year}{2010}).

\bibitem[{\citenamefont{Reese et~al.}(2011)\citenamefont{Reese, Melbinger, and
  Frey}}]{Reese2011}
\bibinfo{author}{\bibfnamefont{L.}~\bibnamefont{Reese}},
  \bibinfo{author}{\bibfnamefont{A.}~\bibnamefont{Melbinger}},
  \bibnamefont{and} \bibinfo{author}{\bibfnamefont{E.}~\bibnamefont{Frey}},
  \bibinfo{journal}{Biophysical Journal} \textbf{\bibinfo{volume}{101}},
  \bibinfo{pages}{2190} (\bibinfo{year}{2011}).

\bibitem[{\citenamefont{Derrida et~al.}(1992)\citenamefont{Derrida, Domany, and
  Mukamel}}]{Derrida1992}
\bibinfo{author}{\bibfnamefont{B.}~\bibnamefont{Derrida}},
  \bibinfo{author}{\bibfnamefont{E.}~\bibnamefont{Domany}}, \bibnamefont{and}
  \bibinfo{author}{\bibfnamefont{D.}~\bibnamefont{Mukamel}},
  \bibinfo{journal}{Journal of Statistical Physics}
  \textbf{\bibinfo{volume}{69}}, \bibinfo{pages}{667} (\bibinfo{year}{1992}).

\bibitem[{\citenamefont{Derrida et~al.}(1993)\citenamefont{Derrida, Janowsky,
  Lebowitz, and Speer}}]{Derrida1993}
\bibinfo{author}{\bibfnamefont{B.}~\bibnamefont{Derrida}},
  \bibinfo{author}{\bibfnamefont{S.~A.} \bibnamefont{Janowsky}},
  \bibinfo{author}{\bibfnamefont{J.~L.} \bibnamefont{Lebowitz}},
  \bibnamefont{and} \bibinfo{author}{\bibfnamefont{E.~R.} \bibnamefont{Speer}},
  \bibinfo{journal}{Journal of Statistical Physics}
  \textbf{\bibinfo{volume}{73}}, \bibinfo{pages}{813} (\bibinfo{year}{1993}).

\bibitem[{\citenamefont{Kolomeisky et~al.}(1998)\citenamefont{Kolomeisky,
  Schutz, Kolomeisky, and Straley}}]{Kolomeisky1998}
\bibinfo{author}{\bibfnamefont{A.}~\bibnamefont{Kolomeisky}},
  \bibinfo{author}{\bibfnamefont{G.}~\bibnamefont{Schutz}},
  \bibinfo{author}{\bibfnamefont{E.}~\bibnamefont{Kolomeisky}},
  \bibnamefont{and} \bibinfo{author}{\bibfnamefont{J.}~\bibnamefont{Straley}},
  \bibinfo{journal}{J. Phys. A: Math. Gen.} \textbf{\bibinfo{volume}{31}},
  \bibinfo{pages}{6911} (\bibinfo{year}{1998}).

\bibitem[{\citenamefont{R\'{a}os et~al.}(2003)\citenamefont{R\'{a}os, Paessens,
  and Sch\"{u}tz}}]{Raos2003}
\bibinfo{author}{\bibfnamefont{A.}~\bibnamefont{R\'{a}os}},
  \bibinfo{author}{\bibfnamefont{M.}~\bibnamefont{Paessens}}, \bibnamefont{and}
  \bibinfo{author}{\bibfnamefont{G.}~\bibnamefont{Sch\"{u}tz}},
  \bibinfo{journal}{Phys. Rev. Lett.} \textbf{\bibinfo{volume}{91}},
  \bibinfo{pages}{238302} (\bibinfo{year}{2003}).

\bibitem[{\citenamefont{M.R.Evans et~al.}(2003)\citenamefont{M.R.Evans,
  R.Juhasz, and L.Santen}}]{Evans2003}
\bibinfo{author}{\bibnamefont{M.R.Evans}},
  \bibinfo{author}{\bibnamefont{R.Juhasz}}, \bibnamefont{and}
  \bibinfo{author}{\bibnamefont{L.Santen}}, \bibinfo{journal}{Physical Review
  E} \textbf{\bibinfo{volume}{68}}, \bibinfo{pages}{026117}
  (\bibinfo{year}{2003}).

\bibitem[{\citenamefont{Reichenbach et~al.}(2006)\citenamefont{Reichenbach,
  Franosch, and Frey}}]{Reichenbach2006}
\bibinfo{author}{\bibfnamefont{T.}~\bibnamefont{Reichenbach}},
  \bibinfo{author}{\bibfnamefont{T.}~\bibnamefont{Franosch}}, \bibnamefont{and}
  \bibinfo{author}{\bibfnamefont{E.}~\bibnamefont{Frey}},
  \bibinfo{journal}{Phys. Rev. Lett.} \textbf{\bibinfo{volume}{97}},
  \bibinfo{pages}{050603} (\bibinfo{year}{2006}).

\bibitem[{\citenamefont{Mukherji}(2007)}]{Sutapa2007}
\bibinfo{author}{\bibfnamefont{S.}~\bibnamefont{Mukherji}},
  \bibinfo{journal}{Phys. Rev. E} \textbf{\bibinfo{volume}{76}},
  \bibinfo{pages}{011127} (\bibinfo{year}{2007}).

\bibitem[{\citenamefont{Klumpp et~al.}(2008)\citenamefont{Klumpp, Chai, and
  Lipowsky}}]{Klumpp2008}
\bibinfo{author}{\bibfnamefont{S.}~\bibnamefont{Klumpp}},
  \bibinfo{author}{\bibfnamefont{Y.}~\bibnamefont{Chai}}, \bibnamefont{and}
  \bibinfo{author}{\bibfnamefont{R.}~\bibnamefont{Lipowsky}},
  \bibinfo{journal}{Phys. Rev. E.} \textbf{\bibinfo{volume}{78}},
  \bibinfo{pages}{041909} (\bibinfo{year}{2008}).

\bibitem[{\citenamefont{Evans et~al.}(2009)\citenamefont{Evans, Ferrari, and
  Mallick}}]{Martin2009}
\bibinfo{author}{\bibfnamefont{M.~R.} \bibnamefont{Evans}},
  \bibinfo{author}{\bibfnamefont{P.~A.} \bibnamefont{Ferrari}},
  \bibnamefont{and} \bibinfo{author}{\bibfnamefont{K.}~\bibnamefont{Mallick}},
  \bibinfo{journal}{J. Stat. Phys.} \textbf{\bibinfo{volume}{135}},
  \bibinfo{pages}{217} (\bibinfo{year}{2009}).

\bibitem[{\citenamefont{Leduc et~al.}(2012)\citenamefont{Leduc, Padberg-Gehle,
  Varga, Helbing, Diez, and Howard}}]{Leduc2012}
\bibinfo{author}{\bibfnamefont{C.}~\bibnamefont{Leduc}},
  \bibinfo{author}{\bibfnamefont{K.}~\bibnamefont{Padberg-Gehle}},
  \bibinfo{author}{\bibfnamefont{V.}~\bibnamefont{Varga}},
  \bibinfo{author}{\bibfnamefont{D.}~\bibnamefont{Helbing}},
  \bibinfo{author}{\bibfnamefont{S.}~\bibnamefont{Diez}}, \bibnamefont{and}
  \bibinfo{author}{\bibfnamefont{J.}~\bibnamefont{Howard}},
  \bibinfo{journal}{Proc. Natl. Acad. Sci. USA} \textbf{\bibinfo{volume}{109}},
  \bibinfo{pages}{6100} (\bibinfo{year}{2012}).

\bibitem[{\citenamefont{Ross}(2012)}]{Ross2012}
\bibinfo{author}{\bibfnamefont{J.~L.} \bibnamefont{Ross}},
  \bibinfo{journal}{Proc. Natl. Acad. Sci. USA} \textbf{\bibinfo{volume}{109}},
  \bibinfo{pages}{5911} (\bibinfo{year}{2012}).

\end{thebibliography}

\begin{widetext}

\begin{table}
  \centering
\caption{Rates $\alpha, \beta, \omega_a, \omega_d$ dependent properties of motor density $\rho$ along MT for MT length $N\to \infty$. Where $\alpha$ is the motor attachment rate to the first site of MT, $\beta$ is motor detachment rate from the last site of MT, $\omega_a, \omega_d$ are the motor attachment and detachment rates to and from MT bulk sites, $K=\omega_a/\omega_d$. $\rho_{l\alpha}$, $\rho_{r\beta}$, and $\rho_r$ are solutions of $(2\rho(x)-1)\rho'(x)=(\Omega_a+\Omega_d)\rho(x)-\Omega_a$ but with boundary conditions $\rho_{l\alpha}(0)=\alpha$, $\rho_{r\beta}(1)=\beta$, and $\rho_r(1)=0.5, \rho_r(0)>0.5$, respectively. For further properties of DW, see \cite{Zhang20101}. The results listed here are only right for $K=\omega_a/\omega_d\ge1$ and $v=1$, results for $K<1$ can be obtained by the motor-empty symmetry, and results for $v\ne1$ can be obtained by replacing $\alpha, \beta, \omega_a, \omega_d$ with $\alpha/v, \beta/v, \omega_a/v, \omega_d/v$ respectively.}
\begin{tabular}{c|c|c|c}
  \hline
  $\alpha$&$\beta$&relations&results\\
  \hline\hline
  \multirow{9}{*}{$0<\alpha<0.5$}& \multirow{3}{*}{$\beta<1/(K+1)$} & $\rho_{r\beta}^{-1}(1-\alpha)\le0$ & BL at $x=0$\\ \cline{3-4}& &
  $\rho_{l\alpha}^{-1}(\beta)\ge1$   & BL at $x=1$ \\ \cline{3-4}& &
  else & DW in (0,1) \\ \cline{2-4}
  &\multirow{3}{*}{$1/(K+1)<\beta<0.5$} &
  $\rho_{r\beta}^{-1}(1-\alpha)\ge0$ & BL at $x=0$\\ \cline{3-4}& &
  $\rho_{l\alpha}^{-1}(\beta)\ge1$   & BL at $x=1$ \\ \cline{3-4}& &
  else & DW in (0,1) \\ \cline{2-4}
  &\multirow{3}{*}{$0.5<\beta<1$} &
  $\rho_{r}^{-1}(0)\ge 1-\alpha$ & BL at $x=0, 1$\\ \cline{3-4}& &
  $\rho_{l\alpha}^{-1}(0.5)\le1$, $\rho_{r}^{-1}(0)<1-\alpha$  & DW in (0,1), BL at $x=1$ \\\cline{3-4}& &
  else & BL at $x=1$\\
  \hline\hline

  \multirow{2}{*}{$0.5<\alpha<1$}& $0<\beta<0.5$ &\multicolumn{2}{|c}{BL at $x=0$}\\ \cline{2-4}
  &$0.5<\beta<1$ & \multicolumn{2}{|c}{BL at $x=0, 1$}\\
  \hline\hline
\end{tabular}
  \label{table1}
\end{table}

\newpage

\begin{figure}
\begin{center}
  \includegraphics[width=450pt]{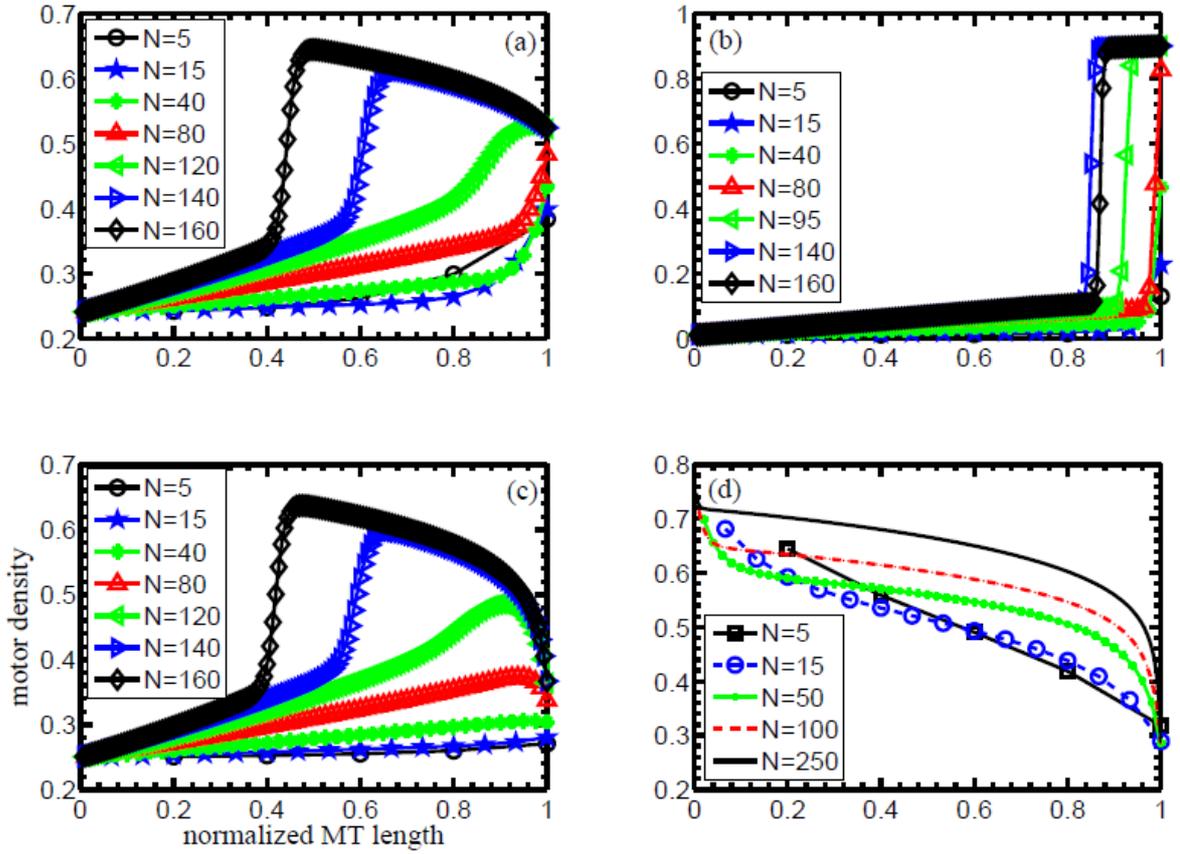}\\
\end{center}
  \caption{Four typical examples of motor density along MT with different length. For the sake of comparison, the MT length is normalized. (a) $\alpha=0.24, \beta=0.48, \omega_a=0.001, \omega_d=0.0002$. Motor traffic is not jammed for short MT, but for MT length $N\ge 120$, motor is jammed from the right boundary. The ratio of jammed MT length $N_H/N$ increases with MT length $N$. (b) $\alpha=0.01, \beta=0.1, \omega_a=0.001, \omega_d=0.0002$. Similarly, motor traffic is not jammed for short MT, but as MT length $N$ increases, motor is jammed from the right boundary. However, the ratio of jammed MT length $N_H/N$ does not change monotonically with MT length $N$. (c) $\alpha=0.25, \beta=0.7, \omega_a=0.001, \omega_d=0.0002$. Similar as in (a), but BL occurs at the MT end for long MT. (d) $\alpha=0.8, \beta=0.9, \omega_a=0.001, \omega_d=0.0008$. There is no DW but BL occurs at the both end of MT for long MT. (a) (b) (c) are examples of $1/(K+1)<\beta<0.5$, $0<\beta<1/(K+1)$, $0.5<\beta<1$ respectively. (d) is an example of $0.5<\alpha<1$ and $0.5<\beta<1$. }\label{FigdensitysampleSKandDW}
\end{figure}

\begin{figure}
\begin{center}
  \includegraphics[width=450pt]{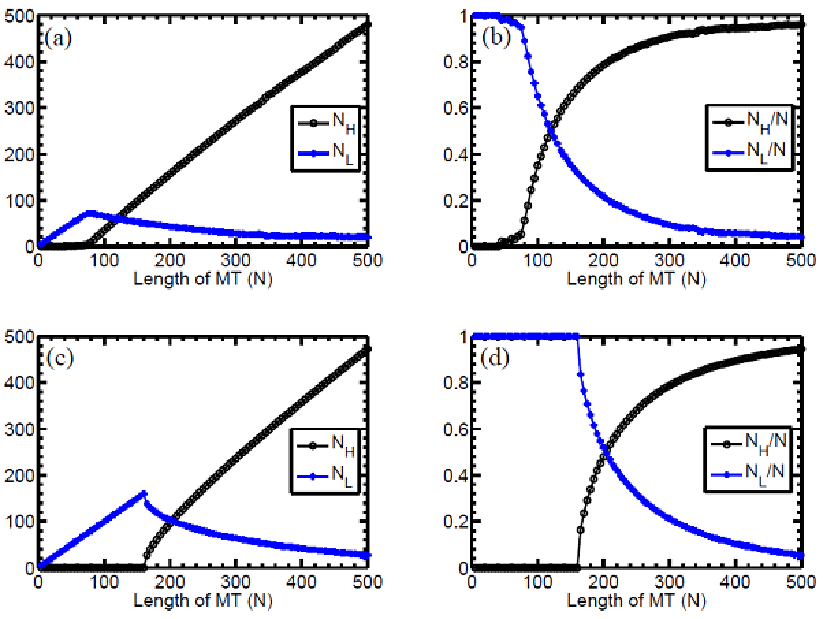}\\
  \includegraphics[width=450pt]{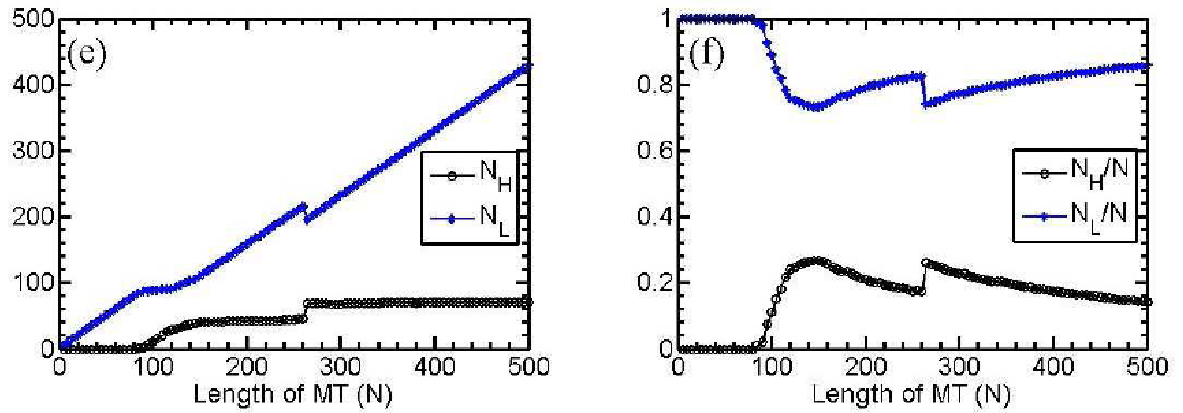}\\
\end{center}
  \caption{Length of MT with high and low motor density, $N_H$ and $N_L$, and their ratios $N_H/N$ and $N_L/N$ for three different cases in which DW occurs for large $N$ limit, see Tab. \ref{table1}. (a, b) $\alpha=0.2, \beta=0.3$, (c, d) $\alpha=0.2, \beta=0.7$, (e, f) $\alpha=0.01, \beta=0.1$. }\label{FigMTdensity}
\end{figure}

\begin{figure}
\begin{center}
  \includegraphics[width=220pt]{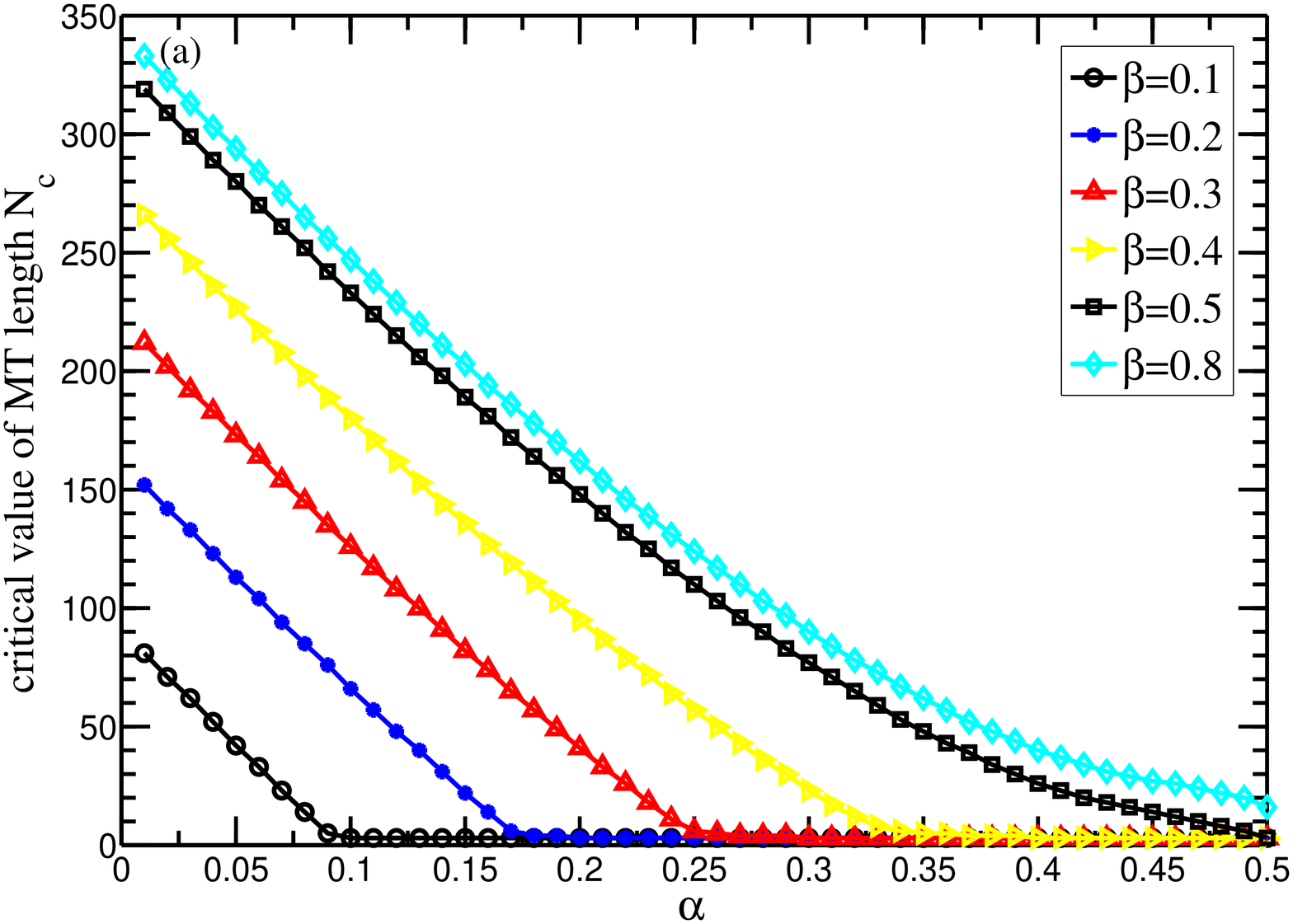}
  \includegraphics[width=220pt]{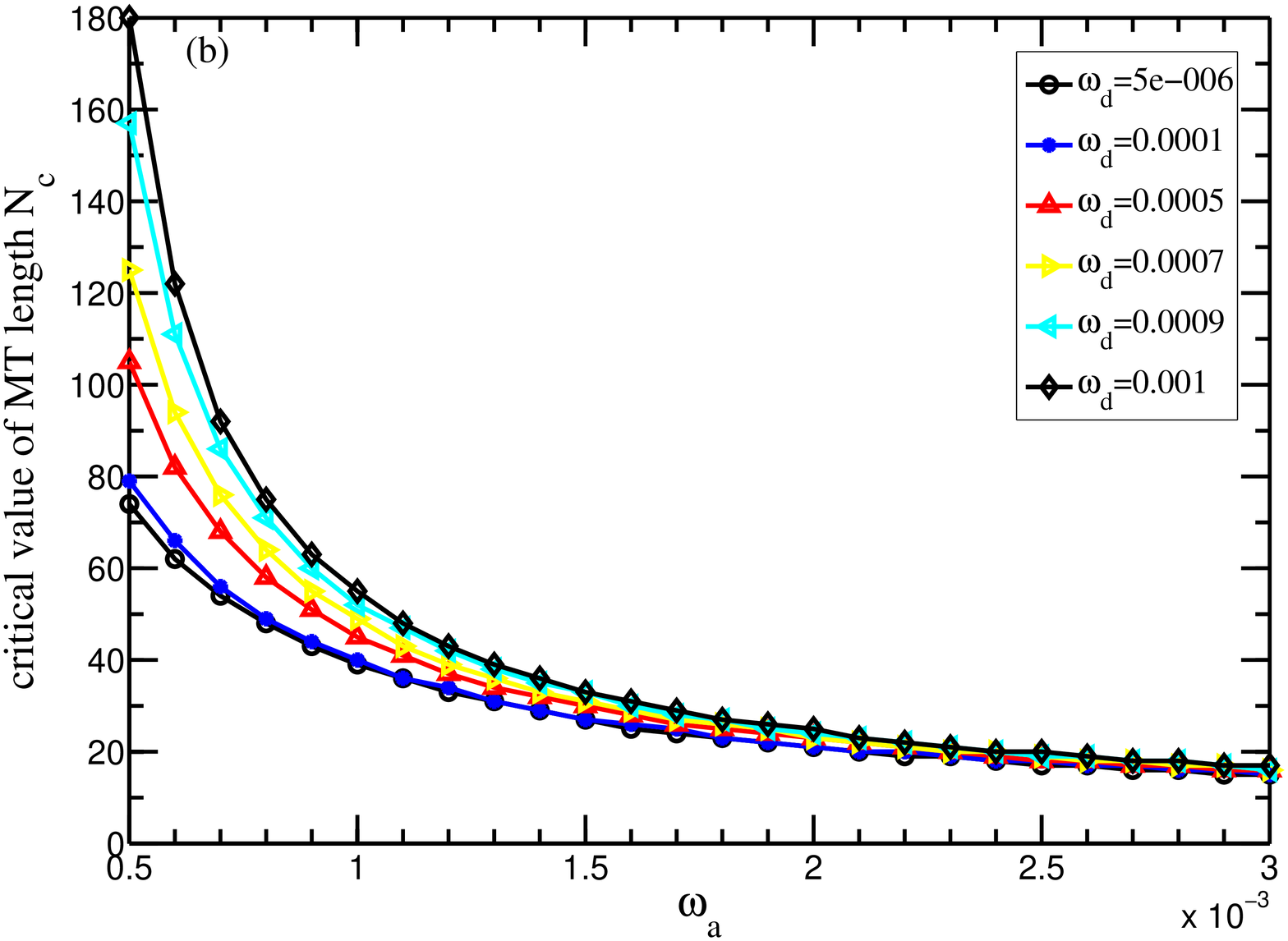}\\
\end{center}
  \caption{The {\it critical value} of MT length $N_c$ decreases with attachment rates $\alpha, \omega_a$, and increases with detachment rates $\beta, \omega_d$. Parameter values used in the calculations are (a) $\omega_a=0.001, \omega_d=0.0002$, (b) $\alpha=0.2, \beta=0.3$. The meaning of {\it critical value} $N_c$ is that, DW occurs iff MT length $N>N_c$.}\label{FigCritical}
\end{figure}

\begin{figure}
\begin{center}
  \includegraphics[width=450pt]{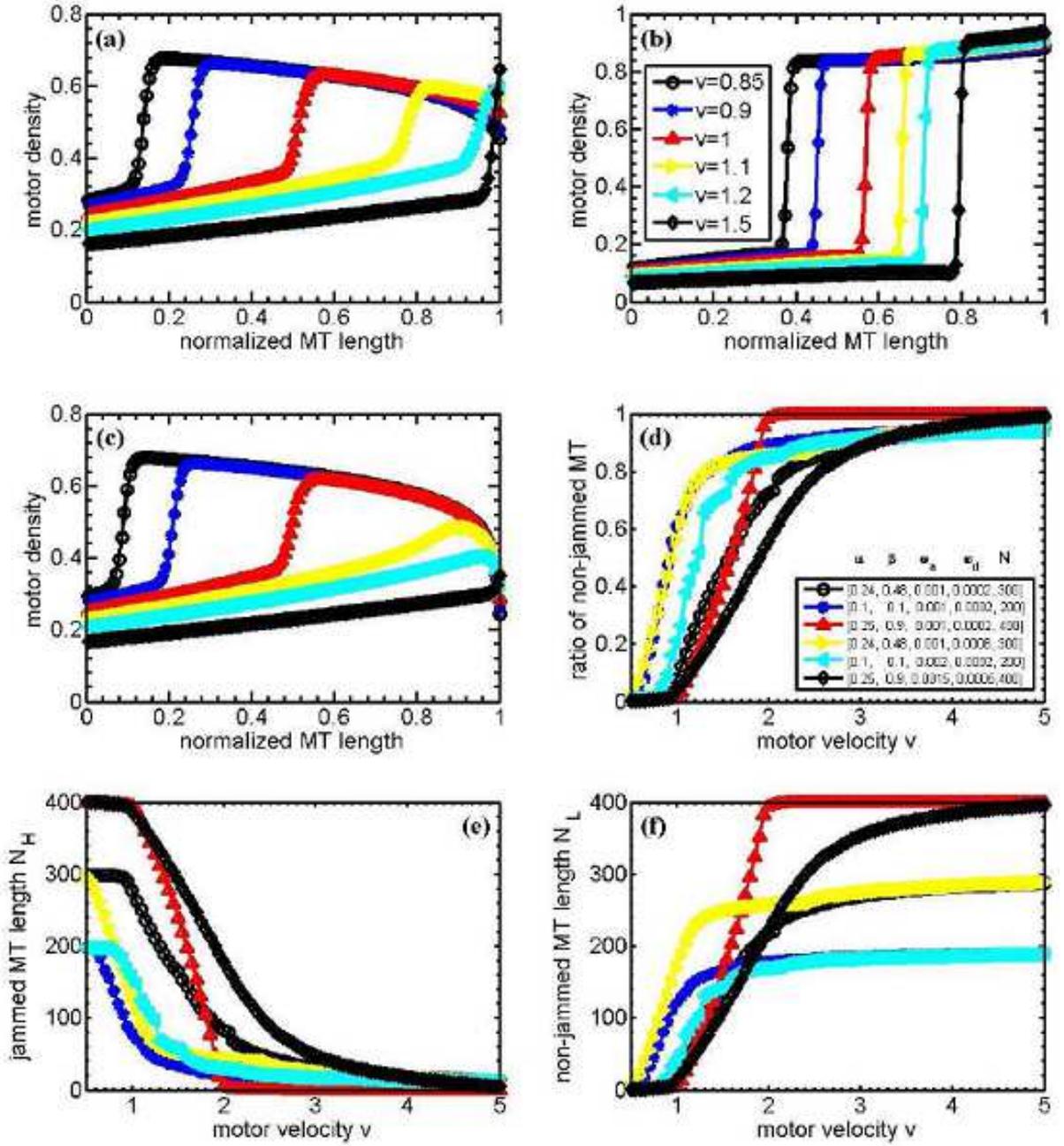}
\end{center}
  \caption{The velocity $v$ dependence of motor traffic along MT. (a, b, c) are examples of steady state motor density for $1/(K+1)<\beta<0.5$, $0<\beta<1/(K+1)$, $0.5<\beta<1$ respectively. The motor velocities used in (a) (c) are the same as given in (b). (d) are ratios of non-jammed MT length $N_L/N$, and (e) (f) are the jammed and non-jammed MT length ($N_H$ and $N_L$) with the same parameters as given in (d). Other parameters used in the calculations of (a, b, c) are $\omega_a=0.001, \omega_d=0.0002, N=150$, and (a) $\alpha=0.24, \beta=0.48$, (b) $\alpha=\beta=0.1$ (c) $\alpha=0.25, \beta=0.9$. The results indicate that $N_L, N_L/N$ increase with motor velocity $v$, and $N_H, N_H/N$ decrease with $v$. }\label{FigPropOfV}
\end{figure}

\end{widetext}

\end{document}